\documentstyle[prl,aps,twocolumn]{revtex}

\begin{document}
\draft

\title{``Interaction--free'' interaction:\\ \
entangling evolution via quantum Zeno effect}

\author{Pawe\l{} Horodecki}

\address{Faculty of Applied Physics and Mathematics\\
Technical University of Gda\'nsk, 80--952 Gda\'nsk, Poland}

\maketitle

\begin{abstract}
The effect of entangling evolution induced
by frequently repeated quantum measurement is
presented. The interesting possibility of
conditional freezing  the system 
in maximally entangled state out of Zeno effect regime 
is also revealed.
The illustration of the phenomena in terms
of dynamical version of
``interaction free'' measurement is presented.
Some general conclusions are provided.
\end{abstract}


Interaction-free measurement (IFM) is one of the fascinating
quantum phenomena.
It has its origin in the
Renninger \cite{Ren} idea of
``negative result experiment''
Then the ``interaction-free'' concept
has been considered by Dicke \cite{IFM1}.
The ideas have been extended and modified
by Elitzur and Vaidman (EV) in their well-known striking
scheme \cite{IFM2} revealing strong modification of interference of
the photon only by {\it presence} of the object.
In their original scheme using one
Mach-Zenhder interferometer in $\eta=25\%$ of events
photon gives us information about presence of
the object - simbolised by ultrasensitive bomb -
with {\it no interaction} with it.
Surprisingly, the  EV scheme
can be improved to obtain the
efficiency $\eta$ to be arbitrarily
close to unity. This result is due to
Kwiat et. al.  \cite{IFM3} who combined
the idea of IFM with the quantum Zeno effect
(QZE) which in another interesting, quantum
phenomenon. In its original form \cite{MS} QZE relies on inhibition
of the decay of unstable quantum system by frequently repeated measurement.
After performance of the proposed experiment \cite{Exp}
there was a discussion  on the essence of the QZE.
While originally it has been explained using quantum collapse,
now it is clear that
it is {\it decoherence} with is the essential ingredient of the effect
\cite{Col1,Decoh,Col2}. Subsequently the effect has been
shown to imply the broad class of different physical phenomena
\cite{AhZen}, \cite{VaidCorr},
\cite{Reson},  \cite{Impos},\cite{Adiab}.

The IFM scheme presented in Ref. \cite{IFM3}
involves two cavities connected via weakly transparent
mirror. This results in a vary small probability
that after one round trip time a photon being initially in left cavity will be found in right one.
But quantum coherent photon evolution
allows for cumulation of the effect of transmission from
left to right (see \cite{IFM3} .
Hence after the time $T$ being sum of large number
of round trip times
the photon will completely leave the left initial cavity
moving to the right one giving finally the probability $p=0$
of finding it in the initial cavity.
Note that the intermediate states of the photon during
the evolution are {\it superpositions}
of two photon states ``present in left cavity'', say, $|0\rangle_{f}$
and ``present in right cavity'', say , $|1\rangle_{f}$.

The situation dramatically changes when one puts the absorbing
object in the right cavity as the evolution of the photon is
{\it no} longer coherent.
After any round trip time $\Delta t$ the superposition
is destroyed - the photom is, in a sense,
forced to decide whether it is in left or right cavity.
This is nothing but quantum-mechanical measurement.
Here the measurement is very frequent if compared with the time $T$
as $\Delta t << T $.
It leads to QZE so that the evolution
corresponding to transition from left to right
is almost stoped
- after time $T$ the photon can be found with probability $p\simeq 1$
in {\it left} cavity. If one consider the value of $p$ in the
absence (presence ) of the object it is clear that testing
the presence of the photon in left cavity
one can almost with certainity
check whether there is an absorbing object in right cavity
although the photon {\it never} interacted with
the object in a sence that it has been neither absorbed
or scattered. The original EV proposal \cite{IFM2} to consider
the object as an ultrasensitive bomb, which explodes if only
one photon is scattered on it, makes the latter effect even more
striking. As showed in  \cite{IFM3}
the presence of the object (bomb) can be simulated by
the atom which can be in one of two levels $|0 \rangle $,
$|1 \rangle $.
The third level $|2 \rangle $ has a very short life time,
after which it decays to the ground state $|0 \rangle $
emitting than
the photon with some frequency $\omega_{signal} $.
Now, if the frequency of the photon $\omega_{f}$ is
adjusted to the transition between states $|1\rangle$, $|2\rangle$
i.e. if one has $\omega_{|1\rangle \rightarrow | 2\rangle} = \omega_{f} $
then the atom in the state $|0 \rangle $ ($|1 \rangle $) corresponds
to the absence (presence) of the bomb in the left cavity.

In this paper we consider the case when the
``quantum bomb'' i.e. the atom  evolves coherently form
state $|0\rangle$ to $|1 \rangle$.
Subsequently we present two effects.
First, we consider the simple dynamical
version of IFM from Ref.\cite{IFM3}.
In this case, as in original proposal,
the photon is almost never absorbed,
so we have not iteraction in this sense.
But we show that now the atom becomes periodically entangled
and disentangled with the photon - so the price for
lack of interaction in one sense (no absorbtion) is payed via
another kind of interaction (entanglement).

Second, we study the case when QZE regime is abondoned.
It can be done in two ways:

(i) One can wait much longer then the
scale of the characteristic time of the photon evolution $T$.
The time $\Delta t$ between measurements is much more
less then $T$ but it is {\it finite}. This implies
that if we wait much longer than $T$ the
approximation $p \simeq 1$ will no longer be true.

(ii) One can make the measurements in the effect {\it not } to
be frequent (then $\Delta t << T$ is false). It would correspond
to the fact that the transparency of the mirror between cavities
is no longer small.

We show that in both cases the system can be
with a great probability {\it conditionally frozed} .
This effect can be described as follows.
Suppose that we have the photon in left cavity,
and the atom in the ground state $| 0\rangle $
i.e. that our system is in the {\it product}
state, and we allow to evolve both of them.
The transmitivity of the mirror between the cavities is not
necessarily small and we are able to find whether
atom absorbed photon detecting then the emitted $\omega_{signal}$.
As we shall see below, it appears, that if we wait for a long time
then with probability  $50 \%$
no detectors will click and our atom + photon system
will stop their evolution being frosen in a maximally entangled state.
We call this effect the conditional forsing,
as it is conditioned by the negative detection
result. But we see that such a frosing
can occur with quite big probability.
At the end of the paper we give simple explanation of the origin
of the effect and formulate some general conjecture.

The photon + atom system will be described using the tensor product
space ${\cal H}={\cal H}_{f} \otimes {\cal H}_{a}$
where the space ${\cal H}_{f}$ (${\cal H}_{a}$)
is spanned by vectors $|0 \rangle_{f}, |1 \rangle_{f} $
($|0 \rangle , |1 \rangle $).
Their tensor products define the product basis in ${\cal H}$.
Any two component system which can be
described in such a way is usually called $2 \times 2$
or two-qubit system and is paradigmatic in investigations
of quantum entanglement.
Sometimes we shall use  maximally entangled states
\begin{eqnarray}
\Psi_{\pm}=\frac{1}{\sqrt{2}}(|0\rangle_{f} |1\rangle
\pm |1\rangle_{f} |0\rangle )
\end{eqnarray}
The evolution of the photon can be
 schematically represented by
\begin{equation}
\phi_{f}(t)=e^{i\sigma_{y} t}\phi_{0}
\label{coherent}
\end{equation}
with $\sigma_{y}$ being second Pauli matrix written
in the basis $\{ |0 \rangle_{f},|1 \rangle_{f} \}$.
Although, in general, the evolution can be more complicated,
the simple picture (\ref{coherent})
allows us to explain the essence of the effects which will take place.
In particular in the above evolution the time
$T$ after which photon being initially in left cavity
(state $|0\rangle_{f}$) with certainity
reaches empty right cavity (state $|1 \rangle_{f} $)
is fixed and simply amounts to $T=\frac{\pi}{2}$.
It is useful here we shall be interested
whether times are small or big {\it relative to time $T$}.

Further we assume the evolution of the atom to be
governed also by the simple Hamiltonian $-\sigma_{y}$.
If we had
$\omega_{|1\rangle \rightarrow | 2\rangle} \neq \omega_{f} $
we would have no measurement and
our system would subject to the product evolution
\begin{eqnarray}
e^{iHt}=e^{i\sigma_{y}t} \otimes e^{-i\sigma_{y}t}, \nonumber \\
\mbox{with} \ \ H=\sigma_{y}\otimes I - I \otimes \sigma_{y}
\label{prod}
\end{eqnarray}
Note that in this case the time $T=\frac{\pi}{2}$
is the moment when both photon and atom meet each other
in the right cavity in the state $|1 \rangle_{f} |1 \rangle $.
In our case this state corresponds to
the ``explosion'' state $\Psi_{exp}=|1 \rangle_{f} |1 \rangle$
becouse we require, as in \cite{IFM3},
$\omega_{|1\rangle \rightarrow | 2\rangle} \neq \omega_{f} $.
Then, if the photon is in the
right cavity (state $|1\rangle_{f}$)
and at the same time atom is in the state
$|1\rangle$, the photon is absorbed
and then after rapid decay the signal photon
is emitted. This effect, simbolising  an ``explosion''
corresponds to the quantum measurement
checking whether the system state is
orthogonal to the product state $|\Psi_{exp}\rangle $.
The corresponding observable is
$P^{\perp}=I - |\Psi_{exp}\rangle \langle \Psi_{exp} | $.
The above measurement will occur after any round trip
time $\Delta t$. Both $\Delta t$ as well as $T$ can be changed by
maneuvering with the size of the cavities, the radiofrequency
driving the atom and the transmitivity of the mirror.
In our picture both times are simply rescaled by fixing $T=\frac{\pi}{2}$.

What will be the evolution of our system if the measurements is
frequent in the sense that $\Delta t << T $ ?

Before we answer the question let us recall some characterisation
of QZE dynamics. It is known (see \cite{Stasz})
that for any bounded Hamiltonian $H$ and projector $P$ we
have the limit
\begin{equation}
lim_{n \rightarrow \infty} (P e^{i H T'/n}P)^{n}=Pe^{iPHPT'}
\label{evol}
\end{equation}
Any initial state $\Psi_{0}$ subjects the above unitary
transformation with
the probability $\langle\Psi_{0}|P|\Psi_{0}\rangle$.
Now one can consider any evolution $e^{iHt}$,
with $H$ bounded, after any $\Delta t$
interrupted by the measurement of observable
$P$. It can be argued that if the measurement is frequent then,
during the time period
$(t_{0}, T)$ with $t_{0} >> \Delta t$,
the formula (\ref{evol}) allows
to approximate such interrupted evolution by
by the {\it new} one taking place
in subspace ${\cal H}'=P{\cal H}$
\begin{equation}
U_{cut}(t)=e^{iH_{cut}t}\mbox{, } \mbox{with } H_{cut}=PHP
\label{ev}
\end{equation}
According to the remark
after formula (\ref{evol}) the above form of dynamics holds for
all initial states $\Psi_{0}$ such that $P^{\perp}\Psi_{0}=\Psi_{0}$.
Summarising - in the limit of very frequent measurements
the system is confined in the subspace $P{\cal H}$
subjecting the drastically different evolution
(\ref{ev}).

Coming back to our photon + atom system we put $P^{\perp}$ in
place of $P$ . We also assume that $T'$ includes
several $T=\frac{\pi}{2}$ periods,
say $T'=5\pi$ or so. As in original IFM scheme
we consider frequent measurement case
which means $\Delta t ={\pi \over 2n} << T \sim T'$.
If initial state of our system belongs to the
subspace ${\cal H}^{\perp}$ spanned by three
vectors $|0 \rangle_{f} |0 \rangle $,  $|0 \rangle_{f} |1 \rangle $,
$|1 \rangle_{f} |0 \rangle$
then the limit of large $n$ the dynamics
is given by the new Hamiltonian
$P^{\perp}(\sigma_{y}\otimes I - I \otimes \sigma_{y})
P^{\perp}$.
It can be easily calculated that it generates the following
limit unitary evolution in subspace ${\cal H}^{\perp}$
\begin{equation}
U_{lim}(t)=
\left [ \begin{array}{ccc}
\cos(\sqrt{2}t) &-\frac{1}{\sqrt{2}}\sin(\sqrt{2}t)  &
\frac{1}{\sqrt{2}}\sin(\sqrt{2}t)  \\
\frac{1}{\sqrt{2}}\sin(\sqrt{2}t)&\cos^{2}(\frac{\sqrt{2}}{2}t)
&\sin^{2}(\frac{\sqrt{2}}{2}t) \\
-\frac{1}{\sqrt{2}}\sin(\sqrt{2}t)&\sin^{2}(\frac{\sqrt{2}}{2}t)&
\cos^{2}(\frac{\sqrt{2}}{2}t)
\end{array} \right ]
\label{limit}
\end{equation}

QZE causes that our system is confined in the subspace
orthogonal to the ``explosion'' state $|1\rangle_{f} |1\rangle $.
Thus, as in original IFM effect
the frequent {\it possibility} of ``explosion''
prevents from {\it actualisation} of it.
But here this possibility, alhough {\it never}
actualised, strongly modifies the evolution of the {\it whole}
system including the atom.
In fact, if the initial state of the system is the product one
$|\psi_{0}\rangle=|0\rangle_{f} |0\rangle $ belonging to
${\cal H}^{\perp}$ then it evolves as
$|\psi(t)\rangle=U_{lim}(t)|\psi_{0}\rangle=\cos(\sqrt{2}t)|0\rangle |0\rangle +
\sin(\sqrt{2}t)\Psi_{-}$.
Thus the system evolves from the product state $\psi_{0}$
to the {\it maximally entangled} state
$\Psi_{-}$ !. So the frequent possibility of ``explosion''
results also in highely {\it entangled} evolution.
The process is ``interaction free''
in one sense ( no absorbtion of the
photon), but the cost for it must
be paid resulting in strong interaction in the other sense
(entanglement).

According to undisturbed product evolution
(\ref{prod}) after time $T=\frac{\pi}{2}$
photon and atom should meet each other in
the ``explosion'' state $|\Psi_{exp} \rangle$.
Here, however, they are in highely entangled state
$|\psi(\frac{\pi}{2})\rangle$ still orthogonal to
$|\Psi_{exp} \rangle$.
It is worth to note that the maximum of entanglement
comes at time $\frac{T}{\sqrt{2}}<T$ hence it happens
before the moment of the supposed meet
at the ``explosion'' state $|\Psi_{exp} \rangle$.


We see that in the process we have in general
two time scales. One of them due to $\Delta t$ describes
the frequency measurements interrupting evolution
or, in other words,  the priods of product evolution.
The another one, represented by the interval
$(t_{0}, T')$, will be called the scale of Zeno
effect. It describes us the region where the Zeno
effect approximation i.e. the formula (\ref{limit}) is good.
What happens, however, if we abandon this scale passing to
the large times $t >> T'$ ?
More precisely, what happens if we keep $\Delta t$ fixed
and take the limit $t \rightarrow \infty $ ?
In general the problem is complicated.
However we can specify our problem as follows.
Assume that we can somehow detect the signal
photon $\omega_{signal}$. No detection up to
time $t$ means that the atom have not
already absorbed the probing photon (the one with
$\omega_{f}$). Let us ask two questions:
(i') is there a nonzero probability that we have never
detected the signal photon up to very large time  ?
(ii') if so, what would be the form of
conditional evolution of our atom + photon system then ?
To answer those questions let us ask about existence of the limit
\begin{equation}
\lim_{n \rightarrow \infty} (P^{\perp}e^{iH\Delta t}
P^{\perp})^{n}=\lim_{n \rightarrow \infty} W(\Delta t)^{n}
\label{as}
\end{equation}
where abbreviation $W(\Delta t)$ has been introduced.
Belowe we shall see that the above limit  exists and posses
property which leads to an interesting effect.

From further analysis we exclude the periods
$\Delta t \neq  k \pi, \frac{k \pi}{2}$
because for them the measurements commutes with
the product evolution (\ref{prod}) having then no impact on it.
Note that those are {\it the only} assumptions
about $\Delta t$. Here the measurement no longer
need to be frequent if compared with $T=\frac{\pi}{2}$.

To see what happens we write it in the new {\it entangled} basis:
\begin{eqnarray}
&&|\Psi_{1} \rangle =\Psi_{+}, \ \ |\Psi_{2}\rangle=
\alpha((2/\tau)|0\rangle_{f} |0\rangle + \sqrt{2}\Psi_{-})  \nonumber \\
&&|\Psi_{3} \rangle=
\beta (-\tau |0\rangle_{f} |0\rangle + \sqrt{2}\Psi_{-}), \ \
|\Psi_{4} \rangle =|1\rangle_{f}|1\rangle
\label{baza}
\end{eqnarray}
with $\tau=\tan (\Delta t)$, $\alpha =\frac{|\tau|}{\sqrt{2\tau^{2}+4}}$,
$\beta=\frac{1}{\sqrt{\tau^{2}+2}}$.
If we restrict to the subspace ${\cal H}'$ then
the limit (\ref{as}) can be written in the form :
\begin{eqnarray}
\lim_{n \rightarrow \infty}{W(\Delta  t)}^{n}=
\lim_{n \rightarrow \infty}
{\left [ \begin{array}{ccc}
1 & 0  &  0 \\
0& -\sin\phi& -\cos\phi \\
0 & \delta \cos\phi &-\delta \sin\phi  \\
0&0&0
 \end{array}
\right ] '}^{n}.
\label{matr}
\end{eqnarray}
Here $\sin\phi=\frac{{\tau}^{2}-2}{{\tau}^{2}+2}$,
$\cos\phi=\frac{4\tau}{\sqrt{2}({\tau}^{2}+2)}$.
The of coefficient $\delta=\cos^{2}(\Delta t)$ is strictly lesser
then 1 for we assumed $\Delta t \neq  k \pi$.
The prime at the matrix in (\ref{matr}) is to stress that it is
written  in the new  {\it entangled} basis (\ref{baza}).

It is no difficult to show that the sequence of $2 \times 2$
submatrix
\begin{equation}
A^{n}=
\left [ \begin{array}{cc}
\sin\phi& \cos\phi \\
\delta \cos\phi &-\delta \sin\phi
 \end{array}
\right ]^{n}
\end{equation}
with $\delta <1 $ tends to zero matrix. To see it
consider the matrix norm $||C||=\mathop{max}\limits_{x}||Cx||$
where maximum is taken over all vectors with norm $||x||=1$.
It has the properties (a) $||AB||\leq||A|| ||B||$ and (b)
$||C^{\dagger}C||=||C||^{2}$.
To show that $\lim_{n \rightarrow \infty } A^{n}=0$
it only suffices to prove that
$\lim_{n \rightarrow \infty} ||A^{2n}||=0$.
From the property (a) of the matrix norm we have
$||A^{2n}||\leq ||A^{2}||^{n}$
and $||A^{2}||=\sqrt{||(A^{T})^{2}A^{2}||}$
Hence we only need to show that the norm of the
new matrix $B=(A^{T})^{2}A^{2}$ is strictly less than
1 i. e. we need to show that $||B||<1$.
The original matrix $A$ can be written as $A=A_{\delta}O$ where
$A_{\delta }=diag(1,-\delta )$ and $O$ is a two-dimensional
rotation around axis x about the angle $\phi+\pi$.
Then (a) we have  $||B||\leq ||A||^{4}\leq
||A_{\delta}||^{2}||O||^{2}=
(max[1,\delta])^{2}\cdot 1=1$ as $O$ is rotation not changing the norm
and the norm of any hermitian operator is equivalent to maximum
of modulus of its eigenvalues.
Thus we have established that $||B||\leq 1$. Note that $B$
is hermitian and has positive eigenvalues as it is of the
form $C^{T}C$. Hence
$||B||=max(b_1,b_2)$ where $b_{1}$,$b_{2}$ are eigenvalues of
B. We already know that none of them is greater then 1.
Can any of them be equal to unity ? The answer is negative.
Indeed, we can calculate explicitly both trace and
determinant of B resulting in $b_{1}b_{2}=detB=\delta^{4}$
and $b_{1}+b_{2}=Tr(B)=\sin^{2}\phi(1+\delta^{4})+2\delta^{2}\cos^2\phi$.
Assumption that some of eigenvalues is equal to 1 gives us,
the equation $(1-\delta^{4})\cos^{2}\phi=0$.
As $\delta<1$ it would imply that
$\cos\phi=0$ but it evidently contaradict
the assumption $\Delta t \neq \frac{\pi}{2}$.
Thus both $b_{1}, b_{2}$ must be less then
unity and then $||B||=max(b_{1},b_{2})<1 $.
According to former remarks it implies
$\lim_{n \rightarrow \infty} A^{n}=0 $.

But what it means for the dynamics of our system ?
The result is quite interesting.
In fact we have
\begin{equation}
\lim_{n \rightarrow \infty} {W(\Delta  t)}^{n}=
diag[1,0,0]'=|\Psi_{+}\rangle \langle\Psi_{+}|
\end{equation}

It means that under the condition
that the negative result of our measurement of
$P^{\perp}$ have not occurred
our system is frozen in maximally entangled state
despite the fact that it has seemingly enough room
to move: three dimensions in the presence of
only one of the orthogonal subspace spanned by
vector $\Psi_{exp}$.

What is the probability of the process ?
It depends on the initial state of the system $\Psi_{0}$.
It can be easily verified that the probability of
staying of the state in the subspace $P^{\perp} {\cal H}$
after $n$ measurements is $p(n\Delta t)=||{W(\Delta  t)}^{n}\Psi_{0}||^{2}$
Taking the limit we obtain simply that
$p(\infty)=|\langle P_{+} | \Psi_{0} \rangle|^{2}$.
It is interesting to examine one
example. Consider our photon plus atom system in the initial
product i.e. {\it disentangled} state $|0\rangle_{f} |0 \rangle $
It means that photon is initially in left cavity and atom is in the
ground state. Let add the detectors to the scheme
waiting for possible signal photon.
Then there is quite large probability
$p(\infty)=\frac{1}{2}=| \langle \Psi_{+ }|0\rangle_{f} |0 \rangle| ^{2}$
that none detector will fire
and that the evolution of our system will gradually stop, frozing finally
photon {\it maximally entangled} with the atom.


The important remark should be given here.
The above process of {\it conditional} frozing the state is
{\it out} of Zeno effect regime also in the sense that,
unlike in original scheme \cite{IFM3}, it does not require
$\Delta t$ to be small. Hence the transmitivity
of the mirror in the present scheme need not be small too.
This effect (as well as the previous one)
could be implemented in other schemes, for example,
those involving QED cavities.

It is illustrative to explain the origin of the
conditional frosing. It is immediate to
see that the surviving state $\Psi_{+}$ is
the only state which is invariant with respect
to both the involved measurement and the product evolution (\ref{prod}).
So the limit of large times simply sweeps out
all the states which are not of this kind.
Following this observation one can expect
that the conditional frosing is quite {\it general}
phenomena occuring in unitary evolution periodically
interrupted by uncomplete von Neumann measurement.
The only important requirements seem to be
the irreducibility of the evolution with respect
to the subspaces defined by the
measurement and the existence of states being
invariants of both evolution and the measurement.

The author thanks M. Czachor, R. Horodecki, 
H. Weinfurter and M. \.Zukowski for helpful discussions
and remarks. He also kindly acknowledges the support
form the Foundation for Polish Science.

\end{document}